\magnification=\magstep1
\hsize=15truecm
\vsize=22.5truecm
\baselineskip 0.55truecm plus 0.01truecm minus 0.01truecm
\parindent=15pt

\font\bfb=cmbx12

\font\rms=cmr9

\def\li{\line}

\def\cl{\centerline}

\def\ni{\noindent}
\def\({\left(}
\def\){\right)}
\def\[{\left[}
\def\]{\right]}

\def\C{\ifmmode \kern1pt^\circ\kern-1.5pt\hbox{C}\else 
$\kern1pt^\circ\kern-1.5pt${C}\fi}

\let\eqln=\eqalignno

\def\bitem#1{\lastdef\bgroup\setbox99=\hbox{#1\enspace}\dimen98=\parindent 
\parindent= \wd99 \setbox97=\hbox{#1}\parskip0pt}
\def\eitem{\par\egroup}
\def\itik#1{\item{\hbox to \wd97{#1\hfil}}}


\input epsf

\overfullrule0pt
\newdimen\rulewidth
\rulewidth0pt 

\def\L{{\rm L}}
\def\R{{\rm R}}
\def\T{{\rm T}}
\def\U{{\rm U}}
\def\D{{\rm D}}
\let\eql=\eqalign
\let\si=\sigma

\def\hr{\noalign{\hrule}}

\def\pstransform#1#2{\hbox to0pt{
#2\hss
}}

\def\Psboxrot#1#2{
\def\epsfsize##1##2{#1##1}%
\setbox0=\vbox{\epsfbox{#2.eps}}%
\dimen1=\wd0
\vbox{\hrule 
height\rulewidth\hbox{\vrule width\rulewidth\vbox to\wd0{\vss\hbox 
to\ht0{\raise\dimen1\pstransform{90 
rotate}{\box0}\hss}}\vrule width\rulewidth}\hrule height\rulewidth}}

\def\spinu{\raise-2pt\hbox to0pt{\hss\tensy\char15\hskip1.5pt\hss}}

\def\spinsd{\raise-2pt\hbox to0pt{\hss\tensy\char15\hskip0.75pt\hss}}
\def\spinsu{\raise15pt\hbox to0pt{\hss\tensy\char15\hskip0.75pt\hss}}

\def\spind{\raise-2.5pt\hbox to0pt{\hss\tensy\char15\hskip1.5pt\hss}}

\def\phfi{\leavevmode\phantom{\fiu}}

\def\fiu
{\leavevmode\hskip1.5pt \vrule height15pt depth0pt width1.5pt\spinu\hskip-0.5pt
\vrule 
height0.5pt depth0pt width15pt\hskip-0.5pt\vrule height15pt depth0pt 
width1.5pt\spinu\hskip1.5pt}

\def\ql
{\leavevmode\smash{\hbox to0pt{\hss\vrule height15pt depth15pt 
width1.5pt\spinu\hskip9pt\hss}}}

\def\qr
{\leavevmode\smash{\hbox to0pt{\hss\hskip9pt\vrule height15pt depth15pt 
width1.5pt\spinu\hss}}}

\def\fid
{\leavevmode \hskip1.5pt \vrule height0pt depth15pt 
width1.5pt\spind\hskip-0.5pt
\vrule height0pt depth0.5pt width15pt\hskip-0.5pt\vrule height0pt depth15pt 
width1.5pt\spind\hskip1.5pt}

\def\sq
{\leavevmode\hskip2pt \vrule height17pt depth0pt width0.5pt\spinsd\spinsu
\rlap{\vrule height17.5pt depth-17pt width15pt}%
\smash{\raise5pt\hbox to0pt{\hss\hskip14pt W\hss}}%
\vrule height0.5pt depth0pt width15pt
\vrule height17pt depth0pt width0.5pt\spinsd\spinsu\hskip2pt}

\def\clap#1{\hbox to0pt{\hss#1\hss}}
\def\L{{\rm L}}
\def\R{{\rm R}}

\def\st{\hbox to1cm{\vrule height1cm width0pt\hfil\tensy\char15\hfil}}
\def\nst{\hbox to1cm{\vrule height1cm width0pt\hfil\tensy\char14\hfil}}
\def\est{\hbox to1cm{\vrule height1cm width0pt\hfil}}



\cl{\bfb Effective-field  approximations, including DMRG method,} 
\cl{\bfb for classical inhomogeneous 2D 
spin lattice models}

\bigskip
\cl{\bf Anton \v Surda}

\cl{Institute of Physics, SAS, SK-842 28 Bratislava, Slovakia}

\bigskip\bigskip

A new approach  to derivation of various effective-field approximations for 
lattice spin models in a unique way is presented. It is shown that 
it can give a number of 
methods, including the DMRG method, that can be used to find generally 
inhomogeneous solutions of 2D classical lattice problems. 
A method,  closely related to the DMRG method but without  necessity to 
perform spin 
renormalization, is derived, yielding results practically not different  from 
the DMRG ones.  The matrix-product wave function of Rommer
and \"Ostlund can be constructed from the output of the method.  
 The computational costs of all the derived methods are smaller 
than those of the DMRG.   Most of the results are applicable to the 1D quantum 
systems, as well.

\bigskip

\bigskip
\cl{\bf 1 Introduction}
\bigskip

The rapid development of computational technique is accompanied by an 
increasing interest in numerical treatments of the problems of statistical 
mechanics. Besides Monte Carlo techniques, which simulate finite systems, 
the effective-field methods, which, in principle, treat infinite systems, became 
recently extremely popular.

After traditional effective-field approaches, as molecular-field approximation and 
quasichemical (Bethe)  approximation, the first sophisticated effective-field 
approach was Baxter's Corner-Matrix Method [1], which gave the possibility of 
systematic improvements of the approximation. This was also the feature of the 
cluster 
transfer-matrix method of the author [2], [7--9], which was used for an 
effective 
 description of the incommensurate structures. Finally, Density 
Matrix Renormalization Group (DMRG)  Method for calculation of the ground 
state of 
one-dimensional quantum models  was developed by White [3, 4]. A little later 
it was 
applied also to the 2D classical lattice models by Nishino and its relation to 
the Baxter's method was pointed out [5, 6].

All these works are closely linked to Kramers-Wannier [10]
suggestion of factorizing the wave
function.

The DMRG method started from the real-space renormalization approaches, where, 
calculating the trace of the Hamiltonian or summing up  the partition function,
spin degrees of freedom where reduced by discarding eigenstates of the 
Hamiltonian corresponding to the highest energy level (lowest eigenvalues of 
the transfer matrix). It was shown by White that, instead of the Hamiltonian, 
the density matrix of the target state is more appropriate for this purpose.  
In this paper we show that if the summation of the partition function is done 
in a proper way no further spin reduction is necessary, and the 
renormalization process by discarding the eigenstates of the density matrix
can be eliminated  from the DMRG method  at all.

It is shown that DMRG method is, in fact, not based on repeated  reduction, 
renormalization,  and  addition of new spin variables as in the 
real-space renormalization group methods, but on a subsequent summation of spin
variables in presence of effective fields when calculating the partition function 
(energy of the ground state in quantum case).

The paper is organized as follows: In Section~2 a general method for derivation
of various of effective-field approximation in a unique way is developed. It
is used, in Section~3,  for re-derivation of the cluster transfer-matrix method and DMRG
method,   and derivation a new effective-field method closely related to
DMRG, but working more effectively. In section~4 the links between our and
Rommer-\"Ostlund's [11, 12, 13] approach are shown, and the correct choice 
of the density
matrix  in the case of non-symmetric or non-Hermitian Hamiltonian
of kinetic or quantum models is elucidated. 

\bigskip
\cl{\bf 2  Method}
\bigskip

The main task of statistical treatment of a classical system of spins 
$\sigma _i$ described  by a Hamiltonian $H(\sigma _i)$  is calculation of the 
partition function 
$$
Z=\sum_{\{\sigma _i\}} \exp[-\beta H(\sigma _i)].
$$

For short-range interactions the Boltzmann's weight $\exp[-\beta H(\sigma _i)]$
can be rewritten in the  form of a product of functions ${\cal T}_i$ which 
depend on spin variables of  few rows. The number of rows is given by the 
range of the interactions. 
$$
\exp[-\beta H(\sigma _i)]=\dots
{\cal T}_{i-1}(\Lambda _{i-1}, \dots,\Lambda _{i-1+k})
{\cal T}_{i}(\Lambda _{i}, \dots,\Lambda _{i+k})
{\cal T}_{i+1}(\Lambda _{i+1}, \dots,\Lambda _{i+1+k})\dots, \eqno(1)
$$
where $ \Lambda_i\equiv \dots\sigma _{i,j-1}, \sigma _{i,j}, \sigma 
_{i,j+1}\dots$ are row spin variables and $k$ is  the interaction range. 
Usually 
the functions ${\cal T}$ are written in a matrix form and called 
transfer matrices. 
We shall treat  ${\cal T}$ as a function, but using the common phraseology  
 we shall call it transfer matrix  (T-matrix).

For the case of nearest-neighbour interactions the elements of the
matrix\break 
${\cal T}(\sigma _{i,1},\dots,\sigma _{i,n};\sigma _{i+1,1},\dots,\sigma 
_{i+1,n})$ 
are equal to the values of the function\break
${\cal T}(\sigma _{i,1},\dots,\sigma _{i,n},\sigma _{i+1,1},\dots,\sigma 
_{i+1,n})$

If 
the summation of the Boltzmann's weight 
for a 2D system 
is performed row by 
row, the calculation of the partition function can be reformulated in a series 
of consecutive  steps 
$$
\eql{
& \sum_{\Lambda_{i+k}} 
{\cal T}_i(\Lambda _{i},\Lambda _{i+1} \dots,\Lambda _{i+k-1},\Lambda _{i+k})
\Phi^\U_{i+k}(\Lambda _{i+1}, \dots,\Lambda _{i+k})
= \lambda_i\Phi^\U_{i+k-1}(\Lambda _{i}, \dots,\Lambda _{i+k-1}) \cr
& \sum_{\Lambda_{i}} 
\Phi^\D_{i}(\Lambda _{i}, \dots,\Lambda _{i+k-1})
{\cal T}_i(\Lambda _{i},\Lambda _{i+1} \dots,\Lambda _{i+k-1},\Lambda _{i+k})
= \lambda_i\Phi^\D_{i+1}(\Lambda _{i+1}, \dots,\Lambda _{i+k}) \cr
}\eqno(2)
$$
or the same in the presence of the functions $\Phi$ from the opposite 
half-plane 
$$
\eql{
& \sum_{\Lambda_{i+k}} 
\Phi^\D_{i}{\cal T}_i
\Phi^\U_{i+k}
= \lambda_i\Phi^\D_{i}\Phi^\U_{i+k-1}\cr
& \sum_{\Lambda_{i}} 
\Phi^\D_{i}
{\cal T}_i\Phi^\U_{i+k}
= \lambda_i\Phi^\D_{i+1}\Phi^\U_{i+k}, \cr
}\eqno(\rm 3a)
$$
where $i=1,\dots, N$, and the $\Phi$'s at the left-hand sides of the 
equations 
are results of previous summations. `U' and `D' are for `up' and `down', 
respectively, 
to distinguish from the next step where the summation will be performed in the 
horizontal directions.

Eqns.~(2)  and (3) are equivalent, but in the effective-field treatment they yield
different results, as the functions $\Phi$ provide effective fields only from below 
in the first case
or from below and above in the second one.

The functions $\Phi_i$ are  eigenfunctions of the T-matrices only for the 
homogeneous solutions deeply in the bulk; 
for inhomogeneous ones
 all the functions are spatially dependent and the upper
and lower functions $\Phi^{\U(\D)}_i$ are different from each other (also 
for symmetric T-matrices).
They 
represent Boltzmann's weights of effective fields by which one half of the 
lattice acts onto the other one. Similarly,  $\Phi^\D_{i}{\cal T}_i$ and 
$\Phi^\D_i {\cal T}_i\Phi^\U_{i+k}$ 
can be considered as Boltzmann's weights of one dimensional problems, and, 
assuming that the functions can be factorized, they may be again treated by  
the T-matrix technique; e.g. $\Phi^\D_{i}{\cal T}_i= \prod_j T_{i,j}$, 
where $T_{i,j}=\Phi^\D_{i,j}{\cal T}_{i,j}$.

Let us study a general one-dimensional problem with $n$ row of spins.
Then, the T-matrices and  ``eigenfunctions'' $\Psi$ are 
defined 
on finite clusters for finite-range interaction problems. Further, for the sake
of simplicity, we shall consider only models with interactions within a square
plaquette.  
$$
\eqln{
&\sum_{\{\sigma _{i,j}\}} T(\sigma_{1,j-1},\dots,\sigma_{n,j-1};\sigma_{1,j},
\dots, \sigma_{n,j} ) \Psi_j^\R(\sigma_{1,j},\dots,\sigma_{n,j})
=\lambda _j
\Psi_{j-1}^\R(\sigma_{1,j-1},\dots,\sigma_{n,j-1}) \cr
&\sum_{\{\sigma _{i,j-1}\}}
\Psi_{j-1}^\L(\sigma_{1,j-1},\dots,\sigma_{n,j-1})
T(\sigma_{1,j-1},\dots,\sigma_{n,j-1};\sigma_{1,j},\dots,\sigma_{n,j}) 
=\lambda_j
\Psi_j^\L(\sigma_{1,j},\dots,\sigma_{n,j})\cr
&  & (4)\cr}
$$

Having in mind that we have
 to perform summation in (2) and (3), we try, in an effective way,  to reduce 
the number of rows, $n$, in the one-dimensional model by one. 
     For that reason we double each column and insert between them  new 
columns with $n-1$ spins, as shown in Fig.~1.

\midinsert
\bigskip
\def\ov{$\overbrace{\hbox to 2cm{\hfil}}$}
\li{\hskip2.4cm$j-1$\hskip2.5cm$j$\hskip2.5cm$j+1$\hfil}
\leavevmode\hskip1.2cm\ov\hskip1cm\ov\hskip1cm\ov

\vskip-0.5cm
{\offinterlineskip\leftskip20pt

\leavevmode\st\est\st\st\est\st\st\est\st

\leavevmode\st\nst\st\st\nst\st\st\nst\st

\leavevmode\st\nst\st\st\nst\st\st\nst\st

\leavevmode\st\nst\st\st\nst\st\st\nst\st

\leavevmode\st\nst\st\st\nst\st\st\nst\st

}

\li{\hskip2.3cm\clap{$P^\T_\L$}\hskip1cm\clap{$P_\R$}\hskip1cm\clap{$T$}\hfil}
\bigskip

\cl{Fig.~1. Decorated one-dimensional lattice. {\tensy\char15} -- old spins 
$\sigma_{i,j} $, {\tensy\char14} -- new spins $\pi_{i,j}$.}
\bigskip
\endinsert

The Boltzmann's weight of the whole 1D lattice  instead of
$$
\dots T_{j-1} T_j T_{j+1}\dots
$$ 
would be
$$
\dots T_{j-1} P^\T_{j-1,\L} P_{j-1,\R} T_j P^\T_{j,\L} P_{j,\R} T_{j+1}\dots
 $$
where the functions $P$ are the transfer matrices between the original columns 
of spins, whose number is doubled, and the new ones
$$
\eqalign{
& 
P^\T_{j,\L}\equiv 
P^\T(\sigma_{1,j},\dots,\sigma_{n-1,j},\sigma_{n,j};\pi_{1,j},\dots,\pi
_{n-1,j}  ) \cr 
& P_{j,\R}\equiv 
P(\pi_{1,j},\dots,\pi_{n-1,j}; 
\sigma'_{1,j},\dots,\sigma'_{n-1,j},\sigma'_{n,j})\cr 
& T_j\equiv T(\sigma'_{1,j},\dots,\sigma'_{n-1,j},\sigma'_{n,j}; 
\sigma_{1,j+1},\dots,\sigma_{n-1,j+1},\sigma_{n,j+1})\cr}
$$

The newly introduced spins $\pi$ ({\tensy\char14}) divide the lattice into 
isolated blocks so that it
can be easily summed up 
over the original spins $\si$ and $\si'$ ({\tensy\char15}). 
After the summation the number of new spins is one less per column and the number of spin degrees 
of freedom is reduced, i.e. the right-hand sides of Eqns.~(2, 3) are found.

Since all the statistical properties of the 1D lattice can be determined
from the functions $\Phi_j$,
they
 will remain unchanged 
after introduction of the new spin variables 
if the following equalities are satisfied:

\bigskip
\li{\hfil\vbox{\hrule
\hbox{\vrule
\vbox{\medskip
\hbox{\qquad$P^\T_{j,\L} P_{j,\R}\Psi^\R_j=\Psi^\R_j$\qquad
$\Psi^\L_j P^\T_{j,\L}P_{j,\R} =\Psi^\L_j$\qquad}
\medskip}
\vrule}
\hrule}\hfil\raise9pt\hbox{(5)}}
\bigskip

{All our further considerations are based on this requirement.} 
\medskip

Clearly, the functions $P$ should be constructed from the known values of the 
function $\Psi$.  There are many possibilities how to satisfy (5). 
Here, we discuss two choices of $P$: 

\bigskip

\ni{\bf(i)}
$$
\eql{
 P^\T_{j,\L}(\sigma_{1,j},\dots,\sigma_{n-1,j},\sigma_{n,j};\pi_{1,j},\dots,\pi
_{n-1,j})  
=\ &\Psi_j^\R(\sigma_{1,j},\dots,\sigma_{n,j})(\rho(\si_{1,j},\dots,\si_{n-1,j}
)) ^{-{1\over 2}}\cdot\cr
& \cdot\delta _{\si_{1,j},\pi_{1,j}}\dots\delta _{\si_{n-1,j},\pi_{n-1,j}}\cr}
\eqno(6)
$$
where
$$
\rho(\si_{1,j},\dots,\si_{n-1,j})=\sum_{\si_{n,j}} 
\Psi_j^\L(\si_{1,j},\dots,\si_{n-1,j},\si_{n,j})
\Psi_j^\R(\si_{1,j},\dots,\si_{n-1,j},\si_{n,j})
$$

This choice of $P$ leads to the  cluster transfer-matrix method [2].

\bigskip

\ni{\bf (ii)}
$$
\eql{
&  P^\T_{j,\L}(\sigma_{1,j},\dots,\sigma_{l,j},\sigma_{l+1,j},\dots,
\sigma_{n,j};\pi_{1,j},\dots,\pi
_{n-k,j})  
=\cr
& =\sum_{\{\tau\}}
\Psi_j^\R(\tau_{1,j},\dots,\tau_{l,j}, \si_{l+1,j},\dots\si_{n,j}) 
\hat\rho^{-{1\over 2}}(\tau_{1,j},\dots,\tau_{l,j}; 
\pi_{l+1,j}\dots\pi_{2l,j})\cdot\cr
& \qquad\cdot \delta _{\si_{1,j},\pi_{1,j}}\dots\delta 
_{\si_{l,j},\pi_{l,j}}\cr
& n-k=2l\cr
} 
\eqno(7)
$$
where $\hat\rho^{-{1\over 2}}$ is the square root of the inverse to the matrix 
$$
\eql{
& \hat\rho(\tau_{1,j},\dots,\tau_{l,j}; \pi_{l+1,j}\dots\pi_{2l,j})=\cr
& = 
\sum_{\{\omega \}}
\Psi_j^\L(\tau_{1,j},\dots,\tau_{l,j},\omega _{l+1,j}\dots\omega _{n,j}) 
\Psi_j^\R(\pi_{l+1,j},\dots,\pi_{2l,j},\omega _{l+1,j}\dots\omega _{n,j}).\cr}
$$

\ni$\hat\rho$ is the density matrix corresponding to the functions 
$\Psi_j^{\L(\R)}$.  In the formulas for $P_\R$, in both cases, the indices R 
and L should be interchanged.

The construction of both $P$'s are based on the idea that $\Psi$ in (5)
meets another $\Psi$ from the neighbouring $P$, forms the density function 
$\rho$ or
density matrix $\hat\rho$, which are canceled by the inverse in $P$, and
finally, it replaced by the same $\Psi$ from the second $P$ in (5).

\medskip
The choice {(ii)} of the  function $P$  reduces the number of spins by $k=n 
-2l$.\break
$k$ can be equal to 1 only if $n$ is odd.

If $k=0$, the choice (ii) gives 
 the Density Matrix Renormalization Group (DMRG)  method. Now, no 
reduction of spin variables takes place and it must be done artificially
 by discarding the spin degrees of freedom 
corresponding 
to   one-half of the density matrix eigenvalues (the smallest ones). In this 
case the requirement (5) is  not satisfied for low-order  approximations. 

In the DMRG, one-half of eigenvectors of the density matrix, 
play 
the r\^ole of $P$.
Their sign may be chosen  arbitrarily, 
what 
sometimes leads to irregularities in the convergence of the iteration process.
The choice (ii) this problems removes. (Note that $P$ corresponds to the 
eigenvectors of the density matrix, which, in distinction to Eqn.~7, is summed 
up over first half of the spins in the argument of $\Psi$'s.)

\bigskip
\cl{\bf 3 Some methods derived from the requirement (5)}
\bigskip

\ni{\bf Cluster transfer-matrix method}
\medskip
The choice (i) of $P$ can be successfully applied to the Eqns.~(2) if the 
possibility of factorization of the functions 
$\Phi_{i+k}^\U=\prod_j \Phi^\U_{i+k,j}(\sigma_{i+1,j},\dots,\sigma_{i+k,j};
\sigma _{i+1,j +1} ,\dots,\break\sigma_{i+k,j+1}) $ is assumed.  
From the fact that the interactions are of short range, the same factorization 
follows for the transfer matrix
${\cal T}_i=\prod_j {\cal T}_{i,j}(\sigma_{i,j},\dots,\sigma_{i+k,j};\break
\sigma _{i,j +1} ,\dots,\sigma_{i+k,j+1}) $. Then, the   
left-hand side of (2) can be expressed as a product $\prod_j \Phi^\U_{i+k,j}{\cal 
T}_{i,j}\equiv \prod_j T_{i,j}$.  Inserting $P^\T_\L P_\R$ 
(choice (i)) between every pair of $T$, and summing up over all original   spins 
$\si$, 
we obtain a  function on $n-1$ rows of spins $\pi$, i.e. the function $\lambda 
\Phi^U_{i+k-1}$ 
$$
\eqln{
& \Phi^\U_{i+k-1}=\prod_j \Phi^\U_{i+k-1,j}\cr
&\Phi^\U_{i+k-1,j}(\pi_{i,j},\dots,\pi_{i+k-1,j}; 
\pi_{i,j+1},\dots,\pi_{i+k-1,j+1})= \cr
& \sum_{\sigma_{i+k,j}\atop\sigma_{i+k,j+1}}
\Psi_j^\L(\pi_{i,j},\dots,\pi_{i+k-1,j},\sigma_{i+k,j})
(\rho(\pi_{i,j},\dots,\pi_{i+k-1,j} ))^{-{1\over 2}}\cdot\cr
&\cdot  \Phi^\U_{i+k,j}(\pi_{i+1,j},\dots,\pi_{i+k-1,j},\sigma_{i+k,j};
\pi _{i+1,j +1} ,\dots,\pi_{i+k-1,j+1},\sigma_{i+k,j+1})\cdot\cr  
& \cdot
{\cal T}_{i,j}(\pi_{i,j},\dots,\pi_{i+k-1,j},\sigma_{i+k,j};
\pi _{i,j +1} ,\dots,\pi_{i+k-1,j+1},\sigma_{i+k,j+1})\cdot\cr 
&\cdot \Psi_{j+1}^\R(\pi_{i,j+1},\dots,\pi_{i+k-1,j+1},\sigma _{i +k,j+1})
 \cdot(\rho(\pi_{i,j+1},\dots,\pi_{i+k-1,j+1} ))^{-{1\over 2}}& (8)
\cr}
$$
where the functions  $\Psi$ are calculated from the Eqns.~(4).

The accuracy of the method is given by  the number of rows $k$ that have 
to be 
as large as possible, despite of the fact that, for our 
short-range interactions,  the function $\cal T$ and $\Phi$ may, in principle, 
be defined only on a two and one  row of spins, respectively.

The cluster transfer-matrix method can be used for calculation 
of spatially dependent
properties of inhomogeneous solutions of lattice models. If the 
inhomogeneities are controlled by the boundary conditions, the method `sees'
only three of them, e.g. the lower, right, and left boundaries, as it is based 
on one of  Eqns. (2) and  Eqns. (4). If the boundary conditions are given on 
sides of a rectangle, information from one of the sides is missing, i.e. 
the cluster transfer matrix method can be used only to a somewhat limited class
of inhomogeneous situations.
For 
homogeneous 
solutions all the quantities in (8) are independent of their indices and 
after necessary number of iterations the  bulk value of  $\Phi$ is 
obtained.

The cluster transfer-matrix method  is much less time consuming than the 
 DMRG method as no matrix diagonalization is necessary. Nevertheless,
the value of the critical temperature of the 2D Ising model is only 
slightly worse than the DMRG one for the same size of $\Phi$ as shown in 
Table~1. 

On the other hand,
if the cluster transfer matrix method is applied to Eq. (3), it can describe an
arbitrary inhomogeneous situation, but
the resulting 
critical temperature is for low-order approximations distinctly higher then  
the exact one and converges with increasing $k$ to the exact value rather 
slowly.

The method can be generalized to long-, but finite, range of interactions and 
modified in several ways. This is described in more details in [2], [7-9].

\bigskip
\ni{\bf Modified DMRG method}
\medskip

To derive the standard DMRG method, we have to factorize the function $\Phi$
(for nearest-neighbour interactions defined  on one row of spins),
according to Baxter [1], by a matrix product of matrices 
$\phi_{\si_j, \si_{j+1}}  (\xi_{j};\xi_{j+1})$, indexed by the lattice 
spins to which they belong: $\Phi_i=\sum_{\{\xi\}}\prod_j \phi_{\si_{i,j}, 
\si_{i,j+1}} 
 (\xi_{i,j};\xi_{i,j+1})$. The multi-value variables $\xi$ acquire $m$ values, 
which control the order of the approximation.

 To provide a deeper insight in the further formulas we shall depict the 
resulting equations in a graphical way.

Let us denote  the  matrix
$\phi^\U_{\si_j, \si_{j+1}}  (\xi^\U_{j};\xi^\U_{j+1})$
 graphically by \fiu. The black dots denote the 
spins $\si_j, \si_{j+1}$ and thick vertical lines the multi-value variables
$\xi^\U_{j},\xi^\U_{j+1}$.

Then, $\Phi^\U$ can be expressed as a matrix product
$\dots$ \fiu\fiu\fiu\fiu\fiu $\dots$. A summation is expected over the 
neighbouring thick lines, which describe the same variable $\xi$. Similarly, 
the close-to-each-other dots denote the same spin $\sigma $.


\medskip
Analogously, $\Phi^\D$ is $\dots$ \raise6pt\hbox{\fid\fid\fid\fid\fid} $\dots$

\medskip

Then, the left-hand side of Eq.~(3a) $\Phi^\L_i {\cal T}_i\Phi^\R_{i+1}$ can be 
depicted as 

\break
{\leftskip50pt
\bigskip

\leavevmode\llap{$i+1$\qquad}\fiu\fiu\fiu\fiu\fiu\fiu\fiu\fiu

\vskip-1pt

\sq\sq\sq\sq\sq\sq\sq\sq  \hfill(3b)

\vskip-9pt

\leavevmode\llap{$i$\qquad\quad}\fid\fid\fid\fid\fid\fid\fid\fid

}

\bigskip
\ni where ${\cal T}_i$ is written as a product of plaquette Boltzmann's weights 
$W$: ${\cal T}_i= \prod_j W^i_{j,j+1}$.

Inserting between each two T-matrices
$T=\vcenter{\vbox{\parindent0pt\hsize=1cm\fiu\par\vskip-1pt \sq\par\vskip-9pt 
\fid} }$ functions $P^\T_\L$ and $P_\R$ of the type (ii) constructed from the 
T-matrix eigenfunctions $\Psi^\R$ and $\Psi^\L$, reducing the number of the rows  
and columns  in the density matrix by one-half, summing up over all the spin 
variables $\sigma $ and $\xi$ in the row $i$, the standard DMRG method is 
obtained. Nevertheless, our approach, comparing with the standard one, shows 
more 
clearly how to to construct the density matrix in the case of non-homogeneous 
solution or non-symmetric T-matrix.

In the one-dimensional system (3) even number of spins in each column is used:
2 original spins of the model used in the Boltzmann's weight $W$ and two 
$m$-valued spins $\xi$. To get odd number of spins, we must omit one of the 
original spins, i.e. also the weight $W$. This step would seemingly lead to 
omission of the whole information about the Hamiltonian of the system.  To 
avoid this, we discard all $W$'s but one.

\bigskip
\li{\hskip4.7cm $j$\hskip0.3cm$j$+1 }
\bigskip
{\leftskip40pt

\leavevmode\llap{$i+1$\qquad}\phfi\phfi\phfi\phfi\ql\fiu\fiu\fiu\fiu\fiu
\vskip-1pt

\fiu\fiu\fiu\fiu\sq \hfill(9)

\vskip-9pt
 
\leavevmode\llap{$i$\qquad\quad}\fid\fid\fid\fid\fid\qr\smash{\raise21.5pt
\hbox{\fid\fid\fid\fid}} 

}
\bigskip

In the diagram there appeared two new symbols \qr\quad which, as it will be 
seen, denote $P_\R({i+1,j})$  and  $P^\T_\L({i,j+1})$.

Summing over  sites $k< j$  in the row $i$, we get  
$\Psi^\L_{i,j}(\xi^\D_{i,j},\si_{i,j},\xi^\U_{i,j})$,
and over the site $j+1$ in the row $i$ and all sites in the row $i+1$, 
we get $\Psi^\R_{i,j}(\xi^\D_{i,j},\si_{i,j},\xi^\U_{i,j})$. From this two 
functions  the density matrix is constructed according to the rule (ii), and  
the functions 
$P^\T_\L({i,j})$ and
$P_\R({i,j})$  are calculated. The density matrix is 
generally non-symmetric. 
For the homogeneous phases $\Psi^\R$ and $\Psi^\L$
are the eigenvectors of the transfer matrix
$T=\vcenter{\vbox{\parindent0pt\hsize=1cm\fiu\par\vskip-9pt \fid}}$,  
they are equal to each other,
and the density matrix
 is symmetric. The number of spins in $\Psi$'s is three, and,  
after inserting $P_\R({i,j})$  and  $P^\T_\L({i,j})$ between the matrices at 
the site $(i,j)$, their number will be reduced.

The function $P^\T_\L({i,j})$ should be remembered for the calculation at the 
site 
$i,j-1$ and $P_\R(i,j)$, together with $P^\T_\L({i,j+1})$ from the previous 
step, is used for calculation of the new value of $\phi^\D_{i+1,j; i+1,j+1}$

\bigskip

{\leftskip70pt

\leavevmode\smash{\raise-3.6pt\hbox{\hskip2.7cm\fid}}

\leavevmode\phantom{\ql}\sq\qquad$\longrightarrow$\qquad \hfill(10a)

\vskip-9pt
\ql\fid\qr

}

\bigskip
\ni or
$$
\eql{
& \phi^\D_{\si_{i+1,j}, \si_{i+1,j+1}}  
(\xi^\U_{i,j};\xi^\U_{i,j+1})=\cr
& \sum_{\si_{i,j},\si_{i,j+1}\atop\xi^\D_{i,j},\xi^\D_{i,j+1}}
P_\R(\xi^\D_{i,j},\si_{i,j},\xi^\U_{i,j})
\phi^\D_{\si_{i,j}, \si_{i,j+1}}(\xi^\D_{i,j};\xi^\D_{i,j+1}) 
P^\T_\L(\xi^\D_{i,j+1},\si_{i,j+1},\xi^\U_{i,j+1})\cdot\cr
\noalign{\vskip-0.4cm}
& \hskip2cm\cdot W(\si_{i,j},\si_{i,j+1},\sigma_{i+1,j},\sigma_{i+1,j+1})\cr
\noalign{\medskip}
& \xi^\U_{i,j} \rightarrow \xi^\D_{i+1,j} \quad  \xi^\U_{i,j+1}
\rightarrow \xi^\D_{i+1,j+1}\cr}\eqno(\rm 10b)
$$

The result of this summation is

\bigskip
\li{\hskip3.8cm$j$$-$1\hskip0.3cm $j$\hskip0.3cm$j$+1\hfil }
\bigskip
{\leftskip40pt

\leavevmode\llap{$i+1$\qquad}\phfi\phfi\phfi\phfi\ql\fiu\fiu\fiu\fiu\fiu
\vskip-1pt

\fiu\fiu\fiu\fiu   \hfill(11)

\vskip-9pt
 
\leavevmode\llap{$i$\qquad\quad}\fid\fid\fid\fid\qr\smash{\raise17pt
\hbox{\fid\fid\fid\fid\fid}} 

}
\bigskip

In the next step a new Boltzmann's weight $W$ can be created from
$\phi^\D_{i,j-1; i,j}$  between the columns 
$j-1$ and $j$ by a procedure inverse to (10), and the sweep 
to the left can be continued. The sweep to the right is performed analogously. 

When the steady state (fixed point) is reached,  the sweeps leave  the functions 
$\Psi^\T_\L$ and $\Psi_\R$ unchanged at any site of the lattice, and the 
calculations, in the framework of the approximation given by the number $m$
of values of the variable  $\xi$
in the matrices $\phi$, are exact. The Boltzmann's weights can be created 
between two arbitrary rows so that   the sweeps can be performed in the 
horizontal and vertical 
directions.  Finally, after sufficient number of sweeps,
the values of Boltzmann's 
weights of effective fields  $P_{\L(\R)}$ 
and $\phi^{\U(\D)}$ are obtained at all the sites of the finite lattice, from 
which all interesting  average quantities can be calculated.

The calculation for homogeneous case is very simple. It consists of three
steps iteratively performed until the fixed point is reached.

 1. The ground state eigenfunction $\Psi$ of the transfer matrix
$T=\vcenter{\vbox{\parindent0pt\hsize=1cm\fiu\par\vskip-9pt \fid}}$,  
is calculated. The functions $\Phi \quad \vcenter{\hbox{\fiu}} 
\vcenter{\hbox{\fid}}$  are assumed to be known from the previous
step. 

2.  Knowing $\Psi$,  the function $P$ is found from (7).

3. The new value of $\Phi$ for the step 1 is calculated from Eqn.~(10).

As for the homogeneous case it is no need to distinguish between right, left
upper, and lower functions, the indices  R, L, U, D are ommited. 

In the effective-field methods spontaneous symmetry-breaking occurs, as a
rule. Thus, the critical temperature is given by the onset of non-zero
magnetization. Actually, no calculation of magnetization is needed, e.g. the
difference $\Phi_{\uparrow\uparrow}(\xi,\xi')-\Phi_{\downarrow\downarrow} 
(\xi,\xi')$ also shows if the system is in ferromagnetic or paramagnetic state.   

The described method is much faster than the standard DMRG method, as for the 
same accuracy the functions $\Psi$ acquire only one-half of the values of 
$\Psi$ in DMRG and the density matrix has  one-quarter of elements.
This statement applies for the improved DMRG algorithms [14], as well,
as it can be used for calculation of the superblock $(T)$ wave function also 
in our case.

However,  far faster than the DMRG is the 
cluster transfer-matrix method, where
instead of density matrix $\hat\rho$ a density function $\rho$ is used,   
and the matrix inversion or matrix diagonalization is replaced by division.

For comparison,
the values of critical interaction constants (inverse of the temperatures)
for all the above-mentioned methods in the case of the homogeneous 2D Ising 
model are given in Table~1.

\bigskip
\ni{\bf Table 1.} Critical values of  the interaction constant $K_c$ of the 
Ising model on the 2D square lattice.
\medskip
{\offinterlineskip
\halign to \hsize{\vrule#&\vrule height10pt depth4pt width0pt
#\tabskip4pt plus1fil& 
#\hfil& \vrule#& \hfil#\hfil&\vrule#& #\hfil& \tabskip0pt\vrule#\cr 
\hr
&& Method&& Number of spin-degrees&&\ $K_c$& \cr
&& && of freedom in $\Psi$&& & \cr 
\hr
&& Cluster matrix method applied to Eq. (3)&& $2^{10}=1024$&&   0.4347& \cr
&& (fastest)&&(no diagonalization)&& & \cr
\hr
&& Cluster matrix method applied to Eq. (2)&&  $2^{10}$=1024&&   0.44111& \cr
&& (fastest)&& (no diagonalization)&& & \cr
\hr
&& Our modification of DMRG&& $m=16$\quad $2m^2=512$&&  0.44050&\cr
\hr
&& Our modification of DMRG&& $m=22$\quad $2m^2=968$&&  0.44052&\cr
\hr
&& Standard DMRG&& $m=16$\quad $4m^2=102\rlap4$&&   0.44050& \cr
&& (slowest)&& && & \cr
\hr 
&& Exact value&&  &&  0.44068& \cr
\hr}}

\bigskip

Since the methods are based on an approximate summation 
of the partition function all interesting thermally averaged physical 
quantities can be found as  functions of coordinates. It was found that 
the simplified DMRG without renormalization yields practically the same values 
as standard DMRG also for other relevant physical quantities.

\bigskip
\cl{\bf  4 Discussion and conclusion }
\bigskip

For re-derivation of DMRG and derivation of the modified DMRG, the Baxter
approximate expression in the form of matrix product for the wave function was   
used. Similar factorization was applied by Rommer and \"Ostlund (RO) [11] for
derivation of their variational approach. Nevertheless, the matrices in both
approaches are different. Baxter's matrices are indexed by two site spins
while Rommer and \"Ostlund's only by one spin. They represent the functions
$P$ in our approach. In formula (9) they appear as vertical lines with one
dot. It is easily seen that if the step in (9) is high of several rows 
the wave function 
$\Psi$ in the vertical direction would be a product of RO matrices. 
The step from (9) to (11) may be taken not only as a calculation 
of $\phi^\D_{i+1,j; i+1,j+1}$ from $\phi^\D_{i,j; i,j+1}$
but also as of $P_\L^\T(i,j)$ from  $P_\L^\T(i,j+1)$ or iterative calculation
of RO eigenfunctions in horizontal directions. This procedure would lead to
the minimum of the free energy.
As formation
of arbitrary steps is, in our approach,   exact at fixed point, we see that
we have to obtain  the same results as RO method, if treated exactly
for arbitrary $m$ as it was  done  here. 
Our method and Rommer-\"Ostlund variational approach  are complementary, but  
their results  slightly differs  from the standard DMRG ones, what can be
seen, e.g. from the fact that 
$\sum_{\si,\xi^\D}
P_\R(\xi^\D,\si,\xi^\U)
P^\T_\L(\xi^\D,\si,\xi^{\prime\U})= \delta_{\xi^{\U}\xi^{\prime\U} }$
is satisfied only for two first methods and not for DMRG.

A direct application of RO 
factorization
in the general approach described in Section~2 would lead to necessity of
singular decomposition of matrices, what we want to avoid in this paper.
We established a  connection between  Baxter's and RO factorization.
The possibility of wave-function factorization in Baxter's way
was assumed and RO factorization derived from it.

We claim that our method avoids the renormalization of the spin variables.
Here, the notion of renormalization is taken in the sense of real-space
quantum renormalization group, where the states corresponding to some
eigenvalues of certain matrix are discarded and the spin variables are mixed
together and lose their identity. This is not done in this paper. In both
choices of $P$ (i,ii) many $\delta$-functions appear so that most of the new
and old spins are really identical. However, the new spins
$\pi_{l+1,j}\dots\pi_{2l,j}$ in (7) are different from the old ones and may be
considered in another sense  as renormalized.

The requirement (5) applies  also to one-dimensional quantum and kinetic
problems.
The matrices $P^\T_\L$ and $P_\R$ have to be inserted between $\Psi$ and $H$ 
in the expression $\Psi^+H\Psi$, where $H$ is the Hamiltonian of the superblock
in the White's approach [2]. Now, from (5) it is clear  that, for  
non-hermitian $H$, the left  and right reduction matrices (here 
$P$'s) should be both equal and non-symmetric, or different and symmetric, 
constructed from the left and right eigenvectors, respectively. 
The fact that the standard derivation of DMRG does not give transparent
prescription for constructing the density matrix in non-symmetric case is
seen in [15], where instead  of the correct choice [16, 17]
of the non-symmetric density matrix is not used the second possibility of
right and left density matrix at the left- and right-hand side of the
superblock operator, respectively,
suggested by us, but the incorrect average of the left and right density
matrices, or only the right density matrix, or the density matrix with mixed
terms.

This approach 
enables effective numerical treatment of one-dimensional non-equilibrium 
kinetic models and quantum systems  approximatively solved by analytical 
methods [18].

It was proposed a new method for derivation of different types 
of effective-field 
approximate methods for calculation  of thermal averages of physical 
quantities and thermodynamic functions of inhomogeneous classical 2D spin 
lattices and zero-temperature properties of 1D quantum lattice system.

The effective fields appear after summation over all spins except a square 
plaquette. In (10) they are applied from below, left, and right and are 
represented 
by  $\phi^\D$, $P^\T_\L$, and $P_\R$, respectively. The effective field  from 
above 
affects the calculation indirectly, through $\Psi$, and consequently $P$.
The effective fields should be really perceived in a generalized sense -- 
while 
in the 
cluster transfer matrix method their Boltzmann's weights are positive in the 
DMRG method, they may be negative.

\bigskip
\ni{\bf Acknowledgment} This work was partially supported by the Slovak Grant 
Agency VEGA project 2/4109/97.

I would like to thank the organizers of the Seminar on DMRG method in 
Dresden  and MECO'24 in Wittenberg for the possibility to participate in the 
meetings and in fruitful discussions with I.~Peschel and T.~Nishino.

\bigskip
\bigskip
\cl{\bf REFERENCES}
\bigskip

\item{[1]}R.~J.~Baxter: {\sl J.~Stat. Phys.}  {\bf 19} (1978) 461

\item{[2]}A.~\v Surda: {\sl Phys. Rev.} B {\bf 43} (1991) 908

\item{[3]}S. R.~White: {\sl Phys. Rev.} B {\bf 48} (1993) 10345 

\item{[4]}S. R.~White: {\sl Phys. Rep.}  {\bf 301} (1998) 187

\item{[5]}T.~Nishino: {\sl J.~Phys. Soc. Jpn. \bf 64} (1995) 3598

\item{[6]}T.~Nishino, K.~Okunishi: {\sl J.~Phys. Soc. Jpn. \bf 65} (1996) 891

\item{[7]}I.~Karasov\'a, A.~\v Surda: {\sl J.~Stat. Phys.} {\bf 70} (1993) 675

\item{[8]} P.~Pajersk\'y, A.~\v Surda: {\sl J.~Stat. Phys.} {\bf 76} (1994) 
1475 

\item{[9]} P.~Pajersk\'y, A.~\v Surda: {\sl J.~Phys. A} {\bf 30} (1997) 4187 

\item{[10]} H. A. Kramers,  G. H. Wannier: {\sl Phys. Rev.} {\bf60} (1941)
263

\item{[11]} S.Rommer, S. \"Ostlund: {\sl Phys. Rev. B \bf 55} (1997) 2164

\item{[12]} S. \"Ostlund, S.Rommer: {\sl Phys. Rev. Lett. \bf 75} (1995)
3537

\item{[13]} M. Andersson, M. Boman,  S. \"Ostlund, cond-mat/9810093

\item{[14]} S. R. White: {\sl Phys. Rev. Lett. \bf 77} (1996)
3633

\item{[15]}  E. Carlon, M. Henkel, U. Schollwoeck: cond-mat/9902041

\item{[16]} X. Wang, T. Xiang: {\sl Phys. Rev. B \bf 56}  (1997) 5061

\item{[17]}     N. Shibata: {\sl J. Phys. Soc. Jpn \bf66} (1997), 2221

\item{[18]} E.~Majern\'{\i}kov\'a: {\sl Phys. Rev.} B {\bf 54} (1996) 3272

\bigskip\bigskip\bigskip

\rms
\ni Dedicated to Doc. Eva Majern\'{\i}kov\'a, DrSc. on the anniversary of her 
birthday. 

\bye

\bye